\documentclass[%
 reprint,
superscriptaddress,
 amsmath,amssymb,
 aps,
]{revtex4-2}

\usepackage[usenames,dvipsnames]{xcolor} 
\usepackage{graphicx}
\usepackage{dcolumn}
\usepackage{bm}

\begin{document}

\preprint{APS/123-QED}

\title{Statistical mechanics  of biomolecular condensates via cavity methods}

\author{N. Lauber}
\thanks{These two authors contributed equally}
\affiliation{Biofisika Institute (CSIC,UPV/EHU), Leioa, Spain}
\affiliation{Department of Philosophy, University of the Basque Country, Leioa, Spain}
\author{O. Tichacek}
\thanks{These two authors contributed equally}
\affiliation{Institute of Organic Chemistry and Biochemistry of the Czech Academy of Sciences, Prague, Czech Republic}
\author{R. Bose}
\affiliation{Max Planck Institute for Molecular Cell Biology \& Genetics, Dresden, Germany}
\author{C. Flamm}
\affiliation{Institute for Theoretical Chemistry, University of Vienna, Vienna, Austria}
\author{L. Leuzzi}
\affiliation{Department of Physics, Universita di Roma la Sapienza, Rome, Italy}
\affiliation{Institute of Nanotechnology, Soft and Living Matter Laboratory,
Consiglio Nazionale delle Ricerche (CNR-NANOTEC),  Rome, Italy}
\author{T-Y Dora Tang}
\affiliation{Max Planck Institute of Molecular Cell Biology \& Genetics, Dresden, Germany}
\author{K. Ruiz-Mirazo}
\affiliation{Biofisika Institute(CSIC,UPV/EHU), Leioa, Spain}
\affiliation{Department of Philosophy, University of the Basque Country, Leioa, Spain}
\author{D. De Martino}
\email{daniele.demartino@ehu.eus}
\affiliation{Biofisika Institute(CSIC,UPV/EHU), Leioa, Spain}
\affiliation{Ikerbasque Foundation, Bilbao, Spain}

\begin{abstract}
Physical mechanisms of phase separation in living systems
can play key physiological roles and have recently been the focus of intensive studies.  
The strongly heterogeneous and disordered nature of such phenomena in the biological domain poses difficult modeling challenges that require going beyond  mean field approaches based on postulating a free energy landscape. The alternative pathway we take in this work is to tackle the full statistical mechanics problem of calculating the partition function in these systems, starting  from microscopic interactions, by means of cavity methods. We illustrate the procedure first on the simple binary case, and  we then apply it successfully to ternary systems, in which the naive mean field approximations are proved inadequate. We then demonstrate the  agreement  with lattice model  simulations, to  finally contrast our theory also with  experiments of coacervate formation by associative de-mixing of nucleotides and poly-lysine in aqueous solution.
In this way, different types of
evidence are provided to support   cavity methods as ideal tools for quantitative modeling of biomolecular condensation, giving an optimal balance between the accurate consideration of spatial aspects of the microscopic dynamics and  the fast computational results rooted in their analytical tractability. 
\end{abstract}

\maketitle


\section{Introduction}
The spatial organization of the components of biological cells is a very important aspect of their physiology \cite{harold2003way} and its nature is eminently physical.
For instance, with regard to metabolism, different  processes require in principle different environmental conditions and segregation mechanisms  to ensure an efficient orchestration of cellular functionalities through compartmentalization. Classical, well understood  examples include oxidative phosphorylation and photosynthesis (performed in specialized organelles, mitochondria and chloroplasts, respectively \cite{berg2015stryer}).   
It has been recently proposed that, apart from compartmentalization through lipid membranes,   living systems could deal  with the problem of creating and controlling    microenvironments by means of  the physical mechanism of phase separation, where  liquid mixtures spatially segregate \cite{lifshitz1980landau}.  Examples range from ATP concentration in stress granules to control of gene expression by chromatin condensation \cite{shin2017liquid}, while a better established mechanism is the storage of carbohydrates into starch and/or glycogen \cite{frayn2009metabolic}, avoiding potential osmotic imbalance.
On the flip side, it is well known that wrong formation of biomolecular condensates is the physical correlate of many prion-based pathologies, like mad cow or Alzheimer's disease, for instance.
Additionally, in the field of origins of life the interest in coacervation has been ``rediscovered'' in recent years \cite{tang2014fatty, donau2020active}  as a simple and highly plausible compartmentalization mechanism under prebiotic conditions (as it was actually suggested in the early days of the field \cite{oparin1924proiskhozhdenie}).

One main difference with respect to classical physical and chemical studies on phase separation is the extremely heterogeneous and complex nature of biological components,  with thousands of different species of microscopic units (that can be complex themselves, like polymers) even in a relatively simple bacterium like {\it E. coli} \cite{milo2015cell}. Besides, the specific focus of investigations in life sciences is centered on problems of control, design and inverse modeling. These aspects spurred the wide use of mean field approximations for  theoretical and computational studies, in particular regarding  the extension of the regular solution model \cite{Sear:2003, Jacobs:2017, Jacobs:2021, carugno2021instabilities}.  

In the case of polymer solutions, the classical Flory-Huggins (FH) model can be used to describe the segregative de-mixing with the formation of multiple phases, each enriched in one respective polymer \cite{Flory:1942,Huggins:1942}. This model was later extended by Voorn and Overbeek (VO) for solutions of oppositely charged poly-ions (i.e., charged polymers) which usually display associative de-mixing, with the formation of one phase enriched in multiple poly-ions \cite{Overbeek:1957}. Both models are mean field approximations that build on the interplay between the entropy that drives the mixing of the system and the enthalpy, which results from the interaction energies (resulting from van der Waals or ionic interactions) between the molecules. As such, they do not explicitly deal with the partition function of the system. In recent years the VO model has been criticized in particular for not being sufficient in explaining the phenomenon of complex coacervation, which involves the phase separation of poly-ions and new models have been proposed, some of which state the partition function of the system explicitly \cite{Minton:2020,Lytle:2017,Sing:2020}, but without any attempt to solve it.

In essence, the main common shortcoming of the aforementioned models is that they tend to neglect spatial correlations by recurring to one-factor approximations, akin to the well-known Curie-Weiss (CW) approximation in magnetic systems \cite{cardy1996scaling}. This is known to  lead to difficulties in presence of idiosyncratic, repulsive interactions and frustration, yielding multi-equilibrium. To overcome these difficulties, in the framework of magnetic systems  more refined mean-field approximations were developed, among which the Bethe-Peierls (BP) approximation \cite{bethe1935statistical, peierls1936statistical}, recently reformulated as cavity methods \cite{mezard2003cavity}, or message passing and belief propagation algorithms \cite{yedidia2003understanding}. The latter are considered important standard methods for the statistical physics of spin glasses and disordered systems, with applications that include inference, information theory and resolution of combinatorial optimization problems \cite{mezard2009information}.  

In this work we will  apply  the BP mean field technique to describe self-assembly of biomolecular condensates, focusing more specifically on the problem of reproducing  numerical simulations of a general grand-canonical heterogeneous lattice model.
The manuscript is organized as follows.
First we will introduce the BP approach and compare it with the regular solution model on a simple standard binary system, providing an analytical formula for the spinodal line that increases quantitatively the match with numerical simulations (in comparison with  the classical formula coming from the regular solution model).
Then will a simple ternary system will be considered, to explore a case where the regular solution model is clearly inadequate (i.e., unable to show even a qualitative agreement with numerical simulations) for the case of mixed repulsive and attractive interactions, known to lead to associative de-mixing. We will demonstrate that the BP approach reproduces much better numerical simulations and provides an immediate method to draw phase diagrams with a semi-quantitative controlled match. This supports the usage of BP to classify much more accurately de-mixing phenomena in ternary systems in the interaction coupling space, where we successfully recapitulate the main three modes of phase separation, i.e. i) associative; ii) segregative and iii) counter-ionic de-mixing.  
The latter aspect paves the way for inverse modeling and inference of couplings from experimental results, so our last section will be devoted to the reconstruction of de-mixing phase diagrams from high-throughput data (experiments with poly-lysine and nucleotides in buffer solution). 
We will finally summarize our findings and draw   potential implications stemming from our  work in a conclusion section.

\section{Results}

\begin{figure}[!!!!ht]
\includegraphics[width=0.45\textwidth]{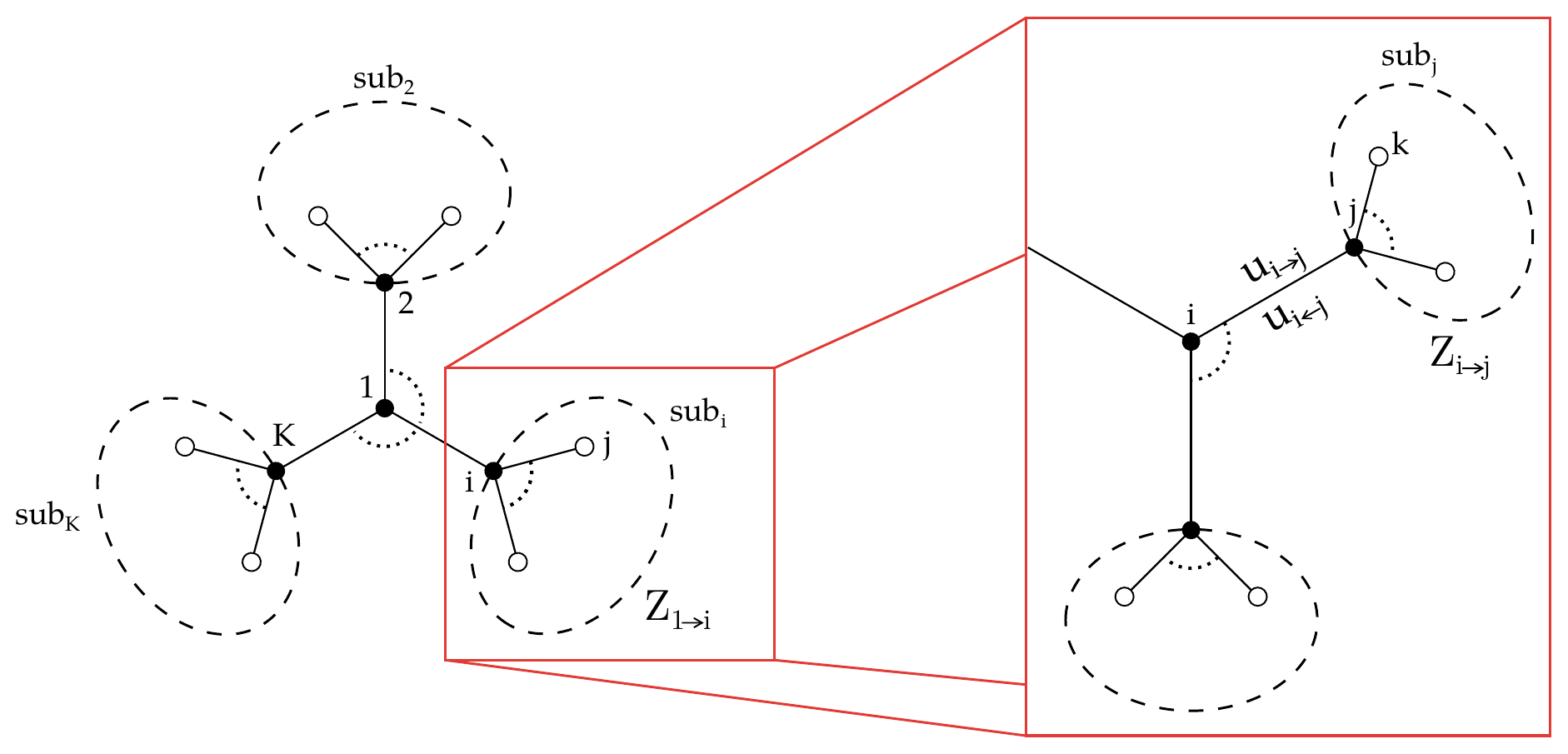}
\caption{Illustration of the tree-like approximation for the calculation of the partition function. Starting from one node, the latter can be decomposed in terms of the conditional partition functions of  subsystems sprouting from neighboring sites and the procedure can be iterated recursively.}
\label{fig:BP-gen}
\end{figure}

As a  starting point, we will consider the microscopic coarse grained multi-component solution model defined  in \cite{Jacobs:2017}, which can be seen as a particular instance of the  Potts model.
The space is discretized into a regular  lattice with $N$ sites, where each site-$i$ is in a state $\sigma_i=0,1\dots q$, standing for the presence of a particle of a given type (e.g., solvent or various solutes).  The interaction between two lattice sites $\sigma_i,\sigma_j$ is described by a given  function $J(\sigma_i,\sigma_j)$ and the number of different kinds of particles is controlled by their chemical potentials  $\mu(\sigma_i)$. The Hamiltonian of the system is thus:
\begin{equation}
  	H(\vec{\sigma}) = -\sum_{\langle i,j\rangle} J(\sigma_i,\sigma_j) - \sum_{i}\mu(\sigma_i)
	  \label{eq:Hamil}
  \end{equation}
where the first sum runs over all neighboring lattice sites $\langle i,j\rangle$. 
In contrast with the regular solution model, we will not postulate a form for the free energy, but rather aim to solve for the partition function.
This is the fundamental quantity that bridges between the molecular microscopic interactions and the collective macroscopic behavior of the system, alongside with its thermodynamic properties \cite{lifshitz1980landau}. Its computation makes it possible to map between the energy as a function of the microscopic configurations and  the free energy as a function of macroscopic variables (e.g. concentrations and/or chemical potentials, temperature).
Here its expression is
\begin{equation}
	  Z = \sum_{\vec{\sigma}}e^{-\beta H(\vec{\sigma})} = \sum_{\sigma_1,\dots,\sigma_N}e^{\beta\left[\sum_{\langle i,j\rangle} J(\sigma_i,\sigma_j) + \sum_{i}\mu(\sigma_i)\right]}
  	\label{eq:PartFun}
\end{equation}

We will compute it by approximating the lattice in terms of a tree-graph branching out from any given site. In this way the lattice  is decomposed in sub-systems that are connected only  by the sprouting site. Once the state value of the latter is fixed, the partition function can be factorized in terms of the partition functions of the subsystems and the procedure can be iterated recursively (see Fig. 1). 
We end up with the equations
\begin{equation}
 Z_{i\rightarrow j}(\sigma_i)=\sum_{\sigma_j}e^{\beta\left[J(\sigma_i,\sigma_j)+\mu(\sigma_j)\right]}\prod_{k\in \partial_{j\setminus i}}Z_{j\rightarrow k}(\sigma_j)
 \label{eq:bpSubPartFun}
 \end{equation}
where $ \partial_{j\setminus i}$ are all the sites connected to $j$ except from $i$,
and $Z_{i\rightarrow j}(\sigma_i)$ is the partition function of the sub-system starting from site-$j$, given that site-$i$ is fixed to the value $\sigma_i$. In general, we have: $Z_{i\rightarrow j}(\sigma_i)\neq Z_{j\rightarrow i}(\sigma_j)$

\subsection{Binary system}
For a simple binary phase separation we have $q=2$ and $\sigma_i=0,1$. Fixing $J(\sigma_i,0)=\mu(0)=0$ the cavity equations will be
\begin{equation}
 Z_{i\rightarrow j}(\sigma_i)=\prod_{k\in N_{j\setminus i}}Z_{j\rightarrow k}(0)+e^{\beta\left[J(\sigma_i,1)+\mu\right]}\prod_{k\in N_{j\setminus i}}Z_{j\rightarrow k}(1)
 \label{eq:bpSubPartFun_bin}
\end{equation}
 Parametrizing $Z_{i\rightarrow j}(\sigma_i)=A_{i\rightarrow j}e^{\beta u_{i \rightarrow j}\sigma_i}$, it is possible to see that equation (4) leads to a set of self-consistent equations for the messages $u_{i\rightarrow j}$:
 \begin{equation}
 u_{i\rightarrow j}=\frac{1}{\beta}\log\left(\frac{1+e^{\beta J+\beta\mu+\beta\sum_{k\in N_{j\setminus i}}u_{j\rightarrow k}}}{1+e^{\beta\mu+\beta\sum_{k\in N_{j\setminus i}}u_{j\rightarrow k}}}\right)
 \label{eq:BinMess}
\end{equation}
Assuming the tree-graph is a Caley-Graph  with a branching of $C=K+1$, and assuming homogeneity $u_{i\rightarrow j} = u,\ \forall i,j$  we get for Eq.\ref{eq:BinMess}:
\begin{equation}
u = \frac{1}{\beta} \log\left[\frac{1 + e^{\beta(J + \mu + Ku)}}{1 + e^{\beta(\mu + Ku)}}\right]
\label{eq:BinMessCal}
\end{equation}
In addition one can assume that the average site occupation or density (equivalent to the occupation probability of the lattice site by the solute) $\langle \sigma \rangle = \phi$  will verify the equation
\begin{eqnarray}
    \phi &= \frac{e^{\beta\mu}\prod_{i\in N_1}Z_{1\rightarrow i}(1)}{\prod_{i\in N_1}Z_{1\rightarrow i}(0) + e^{\beta\mu}\prod_{i\in N_1}Z_{1\rightarrow i}(1)} \\
    &= \frac{e^{\beta\mu+\beta(K+1)u}}{1 + e^{\beta\mu+\beta(K+1)u}}
\label{eq:BinX}
\end{eqnarray}

\begin{figure}[!ht]
\centering
\includegraphics[scale=0.7]{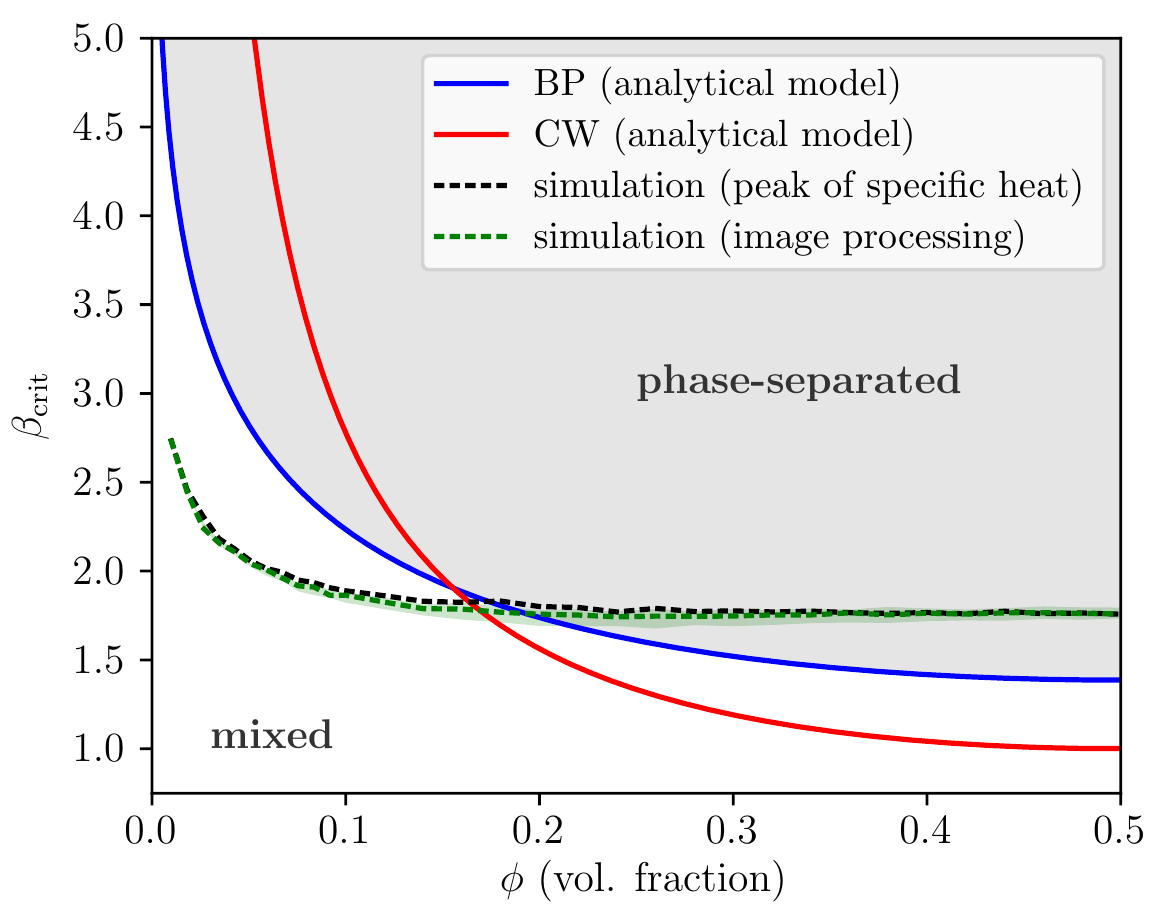}
\caption{Comparison of the critical lines of the mean-field regular solution model (Curie-Weiss-like, red curve), the mean-field finite connectivity cavity method (Bethe-Peierls, blue line) and the numerical simulations on the nearest-neighbor 2D lattice model.  The black line is obtained from the peaks of the specific heat, whereas the green line by means of an image processing method. }
\label{fig:BP-CW}
\end{figure}

These equations express implicitly the state equation $\phi(\mu)$, from which the phase separation curve $(\beta J) (\phi)$ can be obtained, in implicit form, by standard thermodynamic stability analysis upon introducing parameter $w=e^{\beta u}$:  
 \begin{align}
\frac{\phi}{1-\phi} &= \frac{w(w - 1)}{e^{\beta J}-w}, &
w_{1,2} &= \frac{-b \pm\sqrt{b^2 -4K^2e^{\beta J}}}{-2K}, \\
&& b &= Ke^{\beta J} + K - e^{\beta J} + 1
\end{align}
where  the values $w_{1,2}$  correspond to   the two branches of the spinodal line.
The above parametric  formula can be compared now with the one obtained from the regular solution model
\begin{equation}
\beta J =  \frac{1}{(K+1)(1-\phi)\phi}
\end{equation}
and contrasted with microscopic numerical simulations of the  model on a regular square lattice (for which $K=3$) as illustrated in Fig. 2.
As it can be easily observed, even though the model on a nearest-neighbor lattice is extremely far from being tree-like, numerical simulations are in better quantitative agreement with the formula retrieved here by cavity methods, compared with respect to the regular solution equation. 
\subsection{Ternary system}
We next investigated  the phase-separation between two types of solutes and a solvent. We  define configurations as $\sigma_i \in \{-1,0,+1\}$. For the sake of simplicity we assume $J(0,\sigma_i)=J(\sigma_i,0)=0$ $\mu(0)=0$ and rename $J(-1,-1)=J_{--},J(+1,+1)=J_{++}, J(-1,+1)=J(+1,-1)=J_{+-}$ $\mu(-1)=\mu_{-},\mu(+1)=\mu_{+}$.

\begin{figure*}[ht!!!!!!!!!!!!!!!!!]
\centering

\includegraphics[scale=0.85]{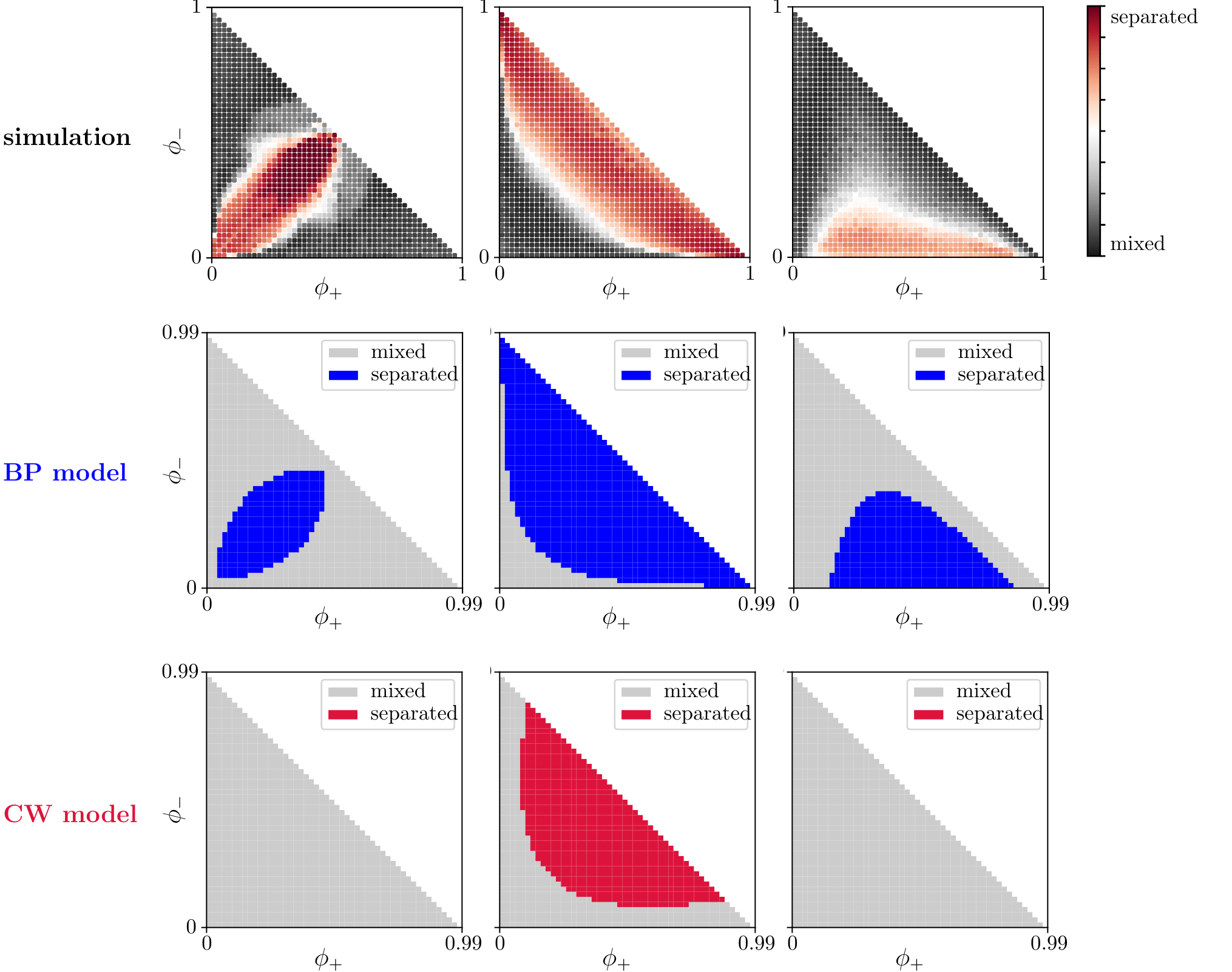}

\caption{Comparison of the phase diagrams in the plane of solutes volume fractions $(\phi_+,\phi_-)$ for the ternary system obtained from lattice model simulations (top), BP mean field (center) and regular solution mean field model (bottom), for the associative (left $J_{--}=J_{++}=-1, J_{+-}=3$), segregative (middle $J_{--}=J_{++}=1, J_{+-}=-3$ ) and counter-ionic (right $J_{++}=2, J_{--}=0, J_{+-}=0.5$) de-mixing cases (see text), respectively.}
\label{fig:tern}
\end{figure*}
In addition, one can apply the BP method to compute the partition function approximately.
In this ternary system the recursive equations for the partition function of subsystems along the branches will be
\begin{equation}
\begin{aligned}
				Z_{i\rightarrow j}(\sigma_i)&=e^{\beta\left[J(\sigma_i,-1)+\mu_{-}\right]}\prod_{k\in \partial_{j\setminus i}}Z_{j\rightarrow k}(-1)\\
			  &+  e^{\beta\left[J(\sigma_i,1)+\mu_{+}\right]}\prod_{k\in \partial_{j\setminus i}}Z_{j\rightarrow k}(+1)
			  \\
			  &+\prod_{k\in \partial_{j\setminus i}}Z_{j\rightarrow k}(0) 
				\end{aligned}
\label{eq:bpSubPartFun_tern}
\end{equation}
Once again these equations can be written in exponential form singling out the dependence on the starting node $\sigma$ in terms of  the so-called message variables.
Restricting ourselves to a homogeneous Cayley tree, we assume homogeneity of the messages $u^{-},u^{+}$, and we obtain  the equations
\begin{equation}
			\begin{aligned}
			u^{-} &= \frac{1}{\beta}\log\left[\frac{e^{\beta(J_{--}+\mu_{-}+Ku^{-})}+1+e^{\beta(J_{+-}+\mu_{+}+Ku^{+})}}{e^{\beta(\mu_{-}+Ku^{-})}+1+e^{\beta(\mu_{+}+Ku^{+})}}\right],\ \\
			u^{+} &= \frac{1}{\beta}\log\left[\frac{e^{\beta(J_{+-}+\mu_{-}+Ku^{-})}+1+e^{\beta(J_{++}+\mu_{+}+Ku^{+})}}{e^{\beta(\mu_{-}+Ku^{-})}+1+e^{\beta(\mu_{+}+Ku^{+})}}\right]
			\end{aligned}
		\label{eq:TernMessCal}
\end{equation}

\begin{figure*}[ht!!!!!!!!!!!!!!]
\centering
\includegraphics[scale=0.75]{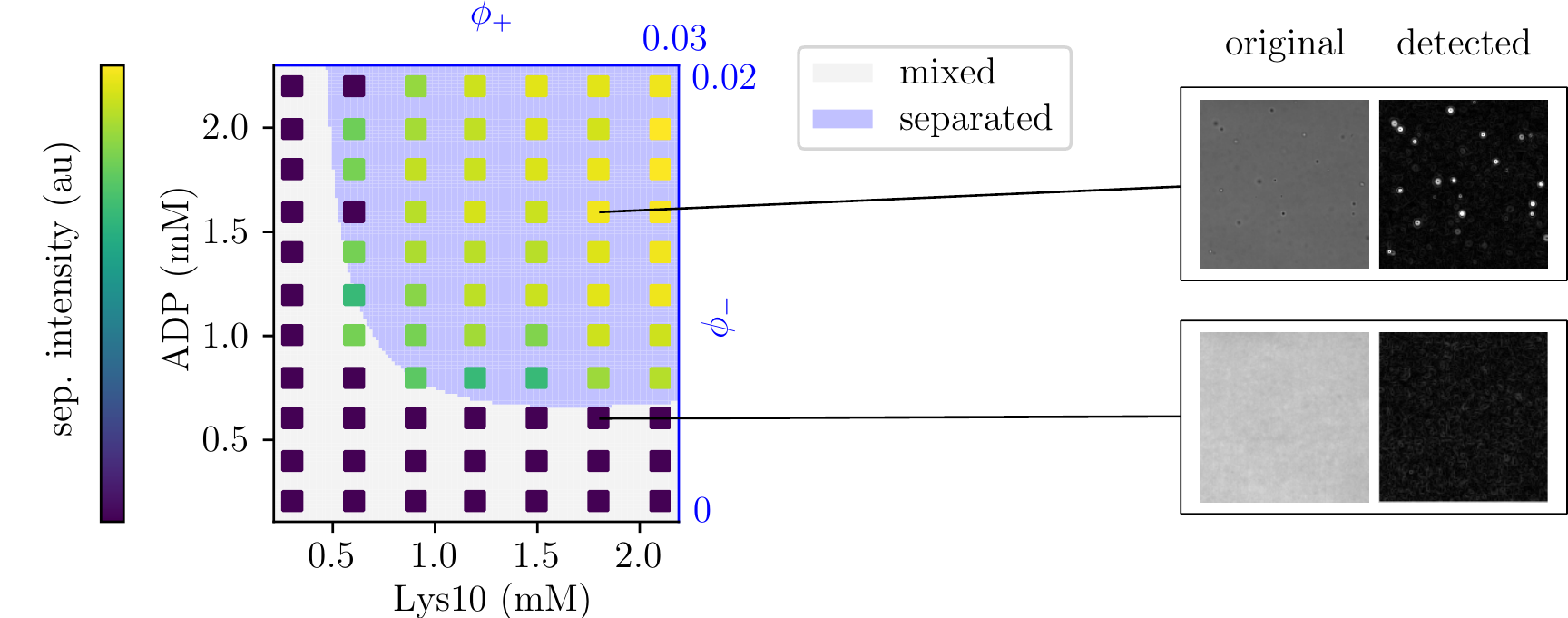}
\caption{Experimental phase diagram in the concentrations plane of a system of poly-lysine and ADP in salt aqueous solution obtained by microscopic imaging, overlaid by a phase diagram of an inferred  ternary system model on a 3D cubic lattice in the volume fractions plane. Example microscope images (right) show the formation of the condensate in the form of droplets. Separation intensity comes from an in-house developed method of automatic image processing of the microscopy images and corresponds to the logarithm of the area of the phase-separated region. The inferred model parameters are $J_{++}=-3.5, J_{--}=-2, J_{-+}=3.8$. Experimental points are reproduced with $97\%$ accuracy.}
\label{fig:tern2}
\end{figure*}      

that together with the average densities 
\begin{equation}
			\begin{aligned}
				\phi_{-} &= \frac{e^{\beta\mu_{-}+\beta(K+1)u^{-}}}{e^{\beta\mu_{-}+\beta(K+1)u^{-}}+1+e^{\beta\mu_{+}+\beta(K+1)u^{+}}}\\[2ex]
				\phi_{+} &= \frac{e^{\beta\mu_{+}+\beta(K+1)u^{+}}}{e^{\beta\mu_{-}+\beta(K+1)u^{-}}+1+e^{\beta\mu_{+}+\beta(K+1)u^{+}}}
			\end{aligned}
			\label{eq:TernProb}
\end{equation}
provide the state equations of the system. The phase diagram can be drawn by checking if the matrix
\[
			H=\begin{pmatrix}
			\frac{\partial\mu_{+}}{\partial \phi_{+}} & \frac{\partial\mu_{+}}{\partial \phi_{-}}\\[2ex]
			\frac{\partial\mu_{-}}{\partial \phi_{+}} & \frac{\partial\mu_{-}}{\partial \phi_{-}}
		\end{pmatrix}
\]
is positive definite.
Results from numerical simulations and mean field calculations are summarized in Fig. \ref{fig:tern}, where depending on the interaction signs phase separation can be classified into three different kinds:
i) associative; ii) segregative and iii) counter-ionic de-mixing. Those three general types of phase behaviour are well accounted for if cavity methods are applied, but more naive or direct mean field models clearly fail to do so.  A strong advantage of mean field approximations is their low computational cost as compared to an explicit lattice model simulation, considering the fact that the system behavior can be assessed by solving a handful of nonlinear equations. In comparative terms, the time to reconstruct the phase diagrams for the ternary system, shown in Fig. \ref{fig:tern}, differs by 6-7 orders of magnitude when we switch from the lattice model simulations to the resolution of the BP equations (hours vs ms in our implementation).
This reduction of computational time paves the way for a full inverse modeling approach to experimental data.
\subsection{Modeling experiments}
We will consider here the experimental phase diagram of a system of poly-lysine and adenosine-diphosphate (ADP) in buffer solution, as obtained by microscopy imaging. 
This system, given the residual electrostatic charge of its components, is expected to show typical associative de-mixing behavior, which is not accounted for correctly by the regular mean field solution model.
We considered thus the task of inferring  the parameters of the aforementioned ternary solution model that reproduces the experimental phase diagram. 
This has been formally modeled as a binary classification problem and we implemented an algorithm for parameter inference based on heuristic optimization via differential evolution algorithms \cite{sarker2002evolutionary}. 
Results are reported in Fig. \ref{fig:tern2}, where we show the experimental phase diagram together with the simulations of the inferred models that are compatible with the associative de-mixing case. Although this provides a quantitative description,  a small perturbation in the initial value of the parameters can lead to an equally well inferred model, with different parameters. 
This hints at the presence of many local maxima for the likelihood of model parameters and calls for more refined experiments and/or an inference scheme going beyond simple binary classification. 
We then performed an inference calculation upon constraining the model parameters to be in the region of segregative de-mixing ($J_{--}>0,J_{++}>0,J_{+-}<0$, not shown),  obtaining a consistently higher error rate (that is the fraction of mismatch in binary phase classification, $15\%$ versus the $3\%$ of the associative case). This shows that our simple setting is able to  tell apart  the different phase separation mechanisms. 
\section{Methods}
\begin{figure*}[ht!!!!!!!!!!!!!!!!!!]

\includegraphics[width=\linewidth]{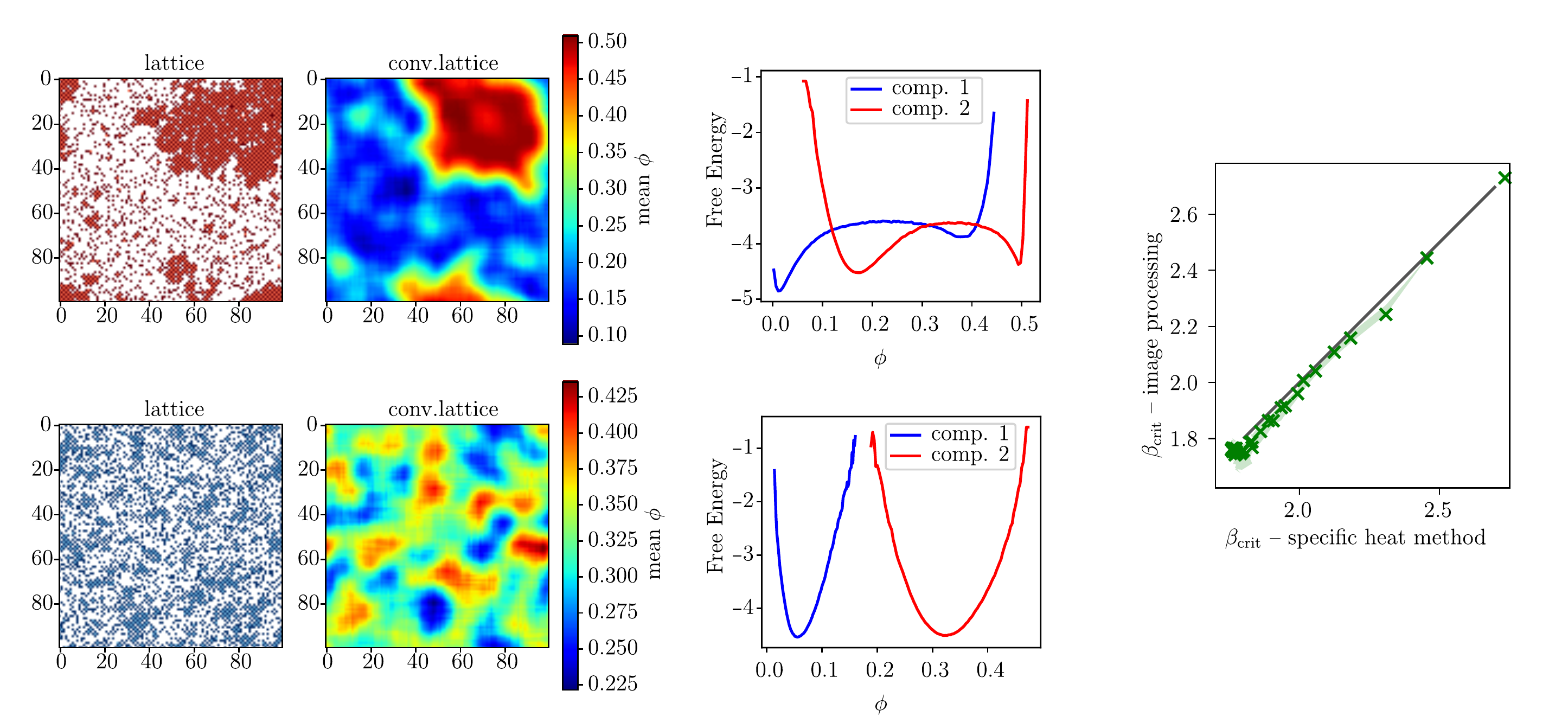}
\caption{Automated detection of phase separation via image processing. Left: lattice model microscopic configuration and its convolution. Center: extracted free energy profile.  Two instances of the ternary systems representative of well mixed and phase separated behavior respectively. A total of 100 snapshots sampled from the end of the simulation were processed independently.
Right: comparison of the inverse critical temperature obtained from specific heat peak computation and image processing automated detection for the binary system.
}
\label{fig:imagproc}
\end{figure*}
\subsection{Numerical methods}
Numerical simulations were performed via the Monte Carlo Kawasaki scheme \cite{kawasaki1966diffusion}, enforcing fixed volume fraction and the inverse critical temperature was estimated independently, as the location of the peak of specific heat and the point where the free energy profile  changes its concavity (see Fig. \ref{fig:imagproc}, right). 
The outcome of Monte Carlo simulations has been analyzed via a heuristic image processing algorithm to identify the occurrence of phase-separation. Local particle densities were computed for each component in a lattice snapshot through 2d convolution with a Gaussian kernel and periodic boundary condition.  The logarithm of the distribution of the local density thus obtained was considered via the Gibbs equation as a bona-fide approximation of the free energy of de-mixing. An automated inspection of the number of minima of this reconstructed profile lead  to the classification of systems into well-mixed and phase-separated. The procedure is inspired by statistical tests comparing nonparametric distributions and produces a separation confidence score depicted as the color scale in the figure \ref{fig:tern}.
 The method is illustrated in Fig. \ref{fig:imagproc}, left, for two instances of the ternary system that are representative of the well mixed and phase separated systems, respectively. The method provided very accurate estimates, as it can be seen in Fig. \ref{fig:imagproc}, right where we show the scattering plot of the inverse critical temperature, at varying volume fractions, for the binary system obtained from the calculation of the peak of the specific heat (through the method described here).
\subsection{Experimental setup}
 Experiments were carried at room temperature in a 1536 well microplate (Greiner bio-one, item no.: 783096) where the coacervate forming components, i.e. Poly-L-lysine hydrochloride of length 10mer (Alamanda polymers, item: PLKC10) and Adenosine 5diphosphate sodium salt (Sigma Aldrich, cat. no.: A2754), were distributed, from 10X stock solutions, using Labcyte Echo 550 acoustic liquid dispenser. The phase diagram was obtained in 20mM Tris-HCl buffer at pH 9.0 which was dispensed using Fluidx XRD-384 reagent dispenser. After mixing the solutions, bright-field images of each of the wells were acquired using Yokogawa CellVoyager\textsuperscript{\texttrademark} CV7000 high-throughput cytological discovery system at 60X magnification (Olympus objective UPLSAPO60XW, product no.: N1480800). The formation of coacervate was detected by visual inspection of the acquired images and by automated detection of phase separation (through image processing).
\section{Conclusions}
The study of how biological cells manage  or fail to control the spatial/physical conditions of their internal milieu through mechanisms of phase separation is of paramount importance. This shall greatly benefit from the wealth of knowledge acquired in  the field of the statistical physics of phase transitions in disordered systems in terms of quantitative modeling, data  analysis and experiment design. In this article we have illustrated an application of the mean field  Bethe-Peierls (BP) approximation in the context of heterogeneous phase separation. We have showed that the BP approach quantitatively  reconstructs phase diagrams where the standard regular solution model fails even to give a qualitative description, more precisely in a minimally heterogeneous ternary lattice microscopic model.
This, apart from being a more adequate theoretical strategy to deal with condensation phenomena, opens a new way for quantitative modeling of experimental data and we provided an example reproducing the experimental phase diagram of the associative de-mixing of poly-lysine in the presence of nucleotides.  The finding that many models lead to a quantitative description of experimental phase diagrams will deserve further investigations.
In this respect, our approach, applied to synthetic data from lattice model simulations could shed light  
in particular with regard to the right experimental quantities to  be measured that lead to well-defined descriptions of the system (i.e., optimal experimental protocols).
In dealing with data an interesting ingredient to analyze with our method would be the introduction of inner degeneracy for the basic degrees of freedom, in order to model complex mixture and that could potentially trigger inverse behaviors \cite{schupper2005inverse}. 
Apart from that, a promising  next step would be to use the BP approach to analyse strongly heterogeneous multi-component systems via replica methods, as it was originally the aim of  the regular solution model. 
In contrast with the latter, BP will be not restricted to the case of mildly attracting interaction matrices. In fact, it could be used to explore any kind of interaction patterns, with idiosyncratic terms, since this approximation showed to be successful in attacking systems with much more complex free energy landscape, like spin glasses. 
\\

\section*{Supplementary material}
Source code for the parameter inference as well as the Monte Carlo numerical simulation is available online: {\small \url{https://github.com/ondrejtichacek/cavity-phase-sep}}.

\begin{acknowledgments}
This work has been supported by the Horizon 2020 Marie Curie ITN (``ProtoMet''—Grant Agreement no. 813873 with the European Commission), within which NL \& RB obtained a PhD fellowship.
OT thanks the Biofisika Institute for kind hospitality during the development of this work.
OT acknowledges the Faculty of Mathematics and Physics of the Charles University (Prague, Czech Republic) where he is enrolled as a PhD student.
RB thanks the Technology Development Studio of MPI-CBG for their technical and infrastructural support. Many thanks to Martin St{\"o}ter setting up the automation of the experimental pipeline and Rico Barsacchi for assisting with the image acquisition.
\end{acknowledgments}




\bibliographystyle{ieeetr}
\bibliography{Ref}

\end{document}